\documentclass[prl,bibnotes,twocolumn,footinbib]{revtex4}
\usepackage{bm}
\usepackage{epsfig}
\usepackage{amsmath}
\usepackage{amssymb}

\newcommand{\vare}{\varepsilon }

\newcommand{\eqb}{\begin{equation}}
\newcommand{\eqe}{\end{equation}}
\newcommand{\pd}{\partial}

\newcommand{\rd}{\mathrm{d}}

\begin{document}

\title{Optical pulse propagation  in a switched-on photonic
lattice: Rabi effect with the  r\^oles of light and matter  interchanged}
\author{V. S. Shchesnovich  }
\affiliation{Centro de Ci\^encias Naturais e Humanas, Universidade Federal do ABC,
 Rua Santa Ad\'elia, 166, 09210-170, \\ Santo Andr\'e, Brazil }

\begin{abstract}
A light pulse propagating in a  suddenly switched on photonic lattice, when the
central frequency lies in the photonic band gap, is an analog of the Rabi model
where the two-level system is the two resonant (i.e. Bragg-coupled) Fourier modes
of the pulse, while the photonic lattice serves as a monochromatic external field.
A simple theory of these  Rabi oscillations is given and confirmed by the numerical
solution of the corresponding Maxwell equations. This is a direct, i.e. temporal,
analog of the Rabi effect, additionally to  the spatial analog in the optical beam
propagation described in \textit{Opt. Lett.}  \textbf{ 32,} 1920 (2007). An
additional high-frequency modulation of the Rabi oscillations reflects the
lattice-induced energy transfer between the electric and magnetic fields of the
pulse.
\end{abstract}

\maketitle

The propagation of  classical waves in periodic structures has been known for a
long time to exhibit intriguing  analogies of the quantum phenomena, such as Bloch
oscillations~\cite{bloch} and  Zener tunneling~\cite{zener}. Optical demonstration
of these two effects have been performed in the one-dimensional periodic structures
based on waveguide arrays and
superlattices~\cite{falk,morand,italy,india,russia,falk2,Zensplat}, and recently in
the two-dimensional case~\cite{2Dexp}. In this letter it is shown that there is yet
another type of analogy in the propagation of optical pulses in the periodic
photonic lattices, where the r\^oles played by the light and matter are
interchanged.

An optical pulse with the central frequency lying inside the photonic band gap is
Bragg reflected by the lattice, which changes its spatial Fourier index by the
reciprocal lattice vector. Applying the same argument to the reflected pulse, we
see that there is a resonant coupling between the two optical modes with the
spatial Fourier indices lying on the opposite sides of the Brillouin zone. The
amplitudes of the resonant Fourier modes are subject to oscillations induced by the
periodic lattice, which can be called Rabi oscillations where the  optical pulse
(its two resonant modes) plays the r\^ole of a two-level system and the photonic
lattice takes the place of an external monochromatic field. Below we provide a
simple theory of this effect and confirm the predictions by direct numerical
simulations.

The spatial analogs of  Rabi oscillations in an optical beam propagating in a
photonic crystal have been previously  studied  \cite{RabiSch,MCPSK} (see also Ref.
\cite{RabiAtom}). The temporal oscillations studied here is a direct analog of Rabi
effect.

\begin{figure}[htb]
\centerline{\includegraphics[width=6cm]{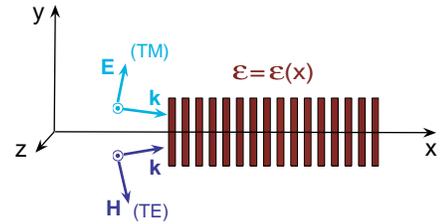}} \caption{(Color online)
Schematic setup. The electric and magnetic fields for the TE-like and TM-like pulse
propagation are shown by the arrows, the photonic lattice is
represented by the bars. }
\label{FG1}
\end{figure}

The optical pulse propagation (oblique, in general, see Fig. \ref{FG1}) is
described by the following equations for the TE-like and TH-like pulses \cite{LL}:
\eqb
\frac{1}{\vare(x)}\nabla^2E = \frac{1}{c^2}\pd^2_t E, \quad
\nabla\frac{1}{\vare(x)}\nabla H = \frac{1}{c^2}\pd^2_t H,
\label{EQ1}\eqe
where $E = \bm{e}_z\bm{E}^{(TE)}$ and $H = \bm{e}_z\bm{H}^{(TM)}$, and $\nabla =
\bm{e}_x\pd_x +\bm{e}_y\pd_y$. Below  we will concentrate on the TE-like pulses,
the TM-like case can be treated similarly.

The refraction index $\vare(x)$ consists of the uniform component $\vare_0$ and a
periodic modulation (a weak photonic lattice), i.e.
\eqb
\frac{1}{\vare(x)} = \frac{1}{\vare_0}\left(1 +  \sum_{n=-\infty}^\infty
\hat{v}_ne^{2ink_Bx} \right),
\label{EQ2}\eqe
where $k_B = \pi/d$ with $d$ being the lattice period and the Fourier amplitudes of
the lattice satisfy $\hat{v}_{-n} = \hat{v}_n^*$ (below, the uniform refraction index
$\vare_0$ is accounted for  by introducing  a modified speed of light $c_0 =
c/\sqrt{\vare_0}$).

\begin{figure}[htb]
\centerline{\includegraphics[width=6cm]{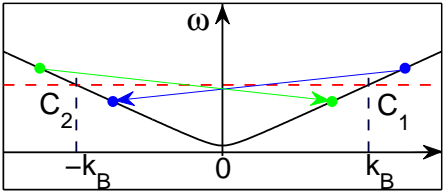}} \caption{(Color online) Bragg
resonant coupling of Fourier modes (shown by dots). The curve gives the dispersion
relation $\omega(k) = c_0\sqrt{k^2+\kappa^2}$ (with $c_0 = 0.9$ and
$\kappa = 0.1$ in dimensionless units). }
\label{FG2}
\end{figure}

Equation (\ref{EQ1}) for the TE-like pulses can be considered in Fourier space by
setting $E(x,y,t) = \Psi(x,t)e^{i\kappa y}$ and using the Fourier transform $\Psi =
\frac{1}{2\pi}\int\rd k C(k,t)e^{ikx}$. We get
\begin{eqnarray}
&&-\frac{1}{c_0^2}\pd_t^2C(k,t) = \left[k^2 +\kappa^2\right]C(k,t) \nonumber\\
&&+ \sum_{n=-\infty \atop n\ne0}^\infty\hat{v}_n\left[(k-2nk_B)^2
+\kappa^2\right]C(k-2nk_B,t).\qquad
\label{EQ3}\end{eqnarray}
Considering the weak lattice limit, $|\hat{v}_1|\ll 1$, the Bragg resonance
condition for $n=1$ is satisfied for the spatial Fourier modes with peaks centered
at $k_B$ and $-k_B$. Assuming the resonant initial pulse $\Psi(x,0)$  with the
Fourier indices $k \in [k_B-\Delta k/2, k_B+\Delta k/2]$ we get
\eqb
-\frac{\rd^2 C_1}{\rd t^2} = \omega_1^2\left[C_1+\hat{v}_1C_2\right], \;
-\frac{\rd^2 C_2}{\rd t^2} = \omega_2^2\left[C_2+\hat{v}^*_1C_1\right],
\label{EQ4}\eqe
where $C_1 = C(k_B+\delta k,t)$ and $C_2 = C(-k_B+\delta k,t)$ are the resonant
Fourier modes, $\omega_{1,2} =  c_0[(k_B\pm \delta k)^2 +\kappa^2]^{\frac12}$ are
the corresponding frequencies, and $\delta k \in [-\Delta k/2, \Delta k/2]$, see
Fig. \ref{FG2}. On the other hand, since Eq. (\ref{EQ4}) is the second-order
system, launching a pulse with the frequencies lying in the forbidden gap is
possible only by suddenly switching on the lattice. The experimental feasibility of
such a setup  is an open question, but possible in principle: the switch-on time is
compared to the Rabi period $T_R \equiv 2\pi/\Omega_0$ with $\Omega_0
=\omega_0|\hat{v}_1|$ (see Eq. (\ref{EQ6}) below) which is much larger than the
light period.

Consider first the simplest case with no detuning, i.e. the modes with the
indices $k_B$ and $-k_B$. We have $\omega(k_B) = \omega(-k_B) = \omega_0$ and the
system (\ref{EQ4})  supplemented with the initial conditions
\eqb
C_1(0) = 1, \; \frac{\rd C_1}{\rd t}(0) = -i\omega_0,\; C_2(0) = 0, \; \frac{\rd
C_1}{\rd t}(0) = 0,
\label{EQ5}\eqe
corresponding to the initially propagating pulse in a homogeneous medium with
$\vare=\vare_0$,  can be easily solved:
\begin{eqnarray}
C_1 = \cos\left(\frac{\Omega_0t}{2}\right)e^{-i\omega_0t} +
\frac{i|\hat{v}_1|}{4}\sin\left(\frac{\Omega_0t}{2}\right)e^{i\omega_0t},\qquad\nonumber\\
C_2 =e^{i\chi}\left[\sin\left(\frac{\Omega_0t}{2}\right)e^{-i\omega_0t}
+\frac{i|\hat{v}_1|}{4}\cos\left(\frac{\Omega_0t}{2}\right)e^{i\omega_0t}\right],
\label{EQ6}\end{eqnarray}
where $\chi = \mathrm{arg}(\hat{v}_1) - \pi/2$ and the terms of order
$O(|\hat{v}_1|^2)$ are omitted. The Fourier powers of the two peaks (denoted here
by $|C_{1,2}|^2$) oscillate with the Rabi frequency $\Omega_0$. The oscillations
are modulated with the  frequency $2\omega_0$ and the amplitude $|\hat{v}_1|/4$.

\begin{figure}[htb]
\centerline{\includegraphics[width=7cm]{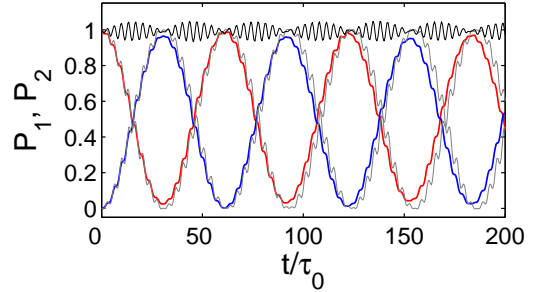}} \caption{(Color online) The
average Fourier powers $P_{1,2}$ (see Eq. (\ref{EQ10}) below) of the numerical
solution of Eq. (\ref{EQ1}), the thick lines, vs the theoretical solution (\ref{EQ6})
with no detuning, the thin  line. The upper line gives the sum of powers $P_1
+P_2$. Here $\tau_0 = (c_0k_B)^{-1}$, $\vare_0 = 1.3$, $\hat{v}_1 = 0.12$, and
$\kappa = 0.05k_B$. The Gaussian initial pulse is used with the width $\sigma= 34d$,
hence $\Delta k = 0.04k_B$, i.e. $\Delta \omega = 0.65\Omega_0$. }
\label{FG3}
\end{figure}

The predictions have been checked by numerical simulations of Eq. (\ref{EQ1}) with the
truncated expression $\vare_\mathrm{num}(x) = \vare_0\left[1 -
\hat{v}_1\cos\left(2\pi x/ d\right)\right]$ (i.e. only the resonant terms are used). The
Gaussian pulse $\Psi(x,0)= \exp\{ik_Bx-(x-\langle x\rangle)^2/\sigma^2\}$ has been used
as the initial condition. Fig. \ref{FG3} shows a good comparison with the
theoretical solution (\ref{EQ6}) for a pulse with the frequency interval $\Delta\omega
\approx 0.07\omega_0 = 0.65\Omega_0$.

\begin{figure}[htb]
\centerline{\includegraphics[width=6cm]{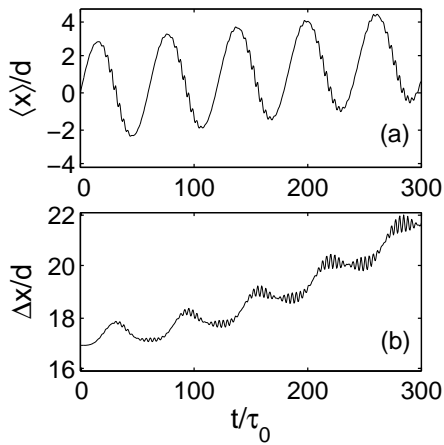}} \caption{The pulse average
position, defined as \mbox{$\langle x\rangle = \int\rd x\, x|E|^2/\int\rd x\,
|E|^2$} panel (a), and its half-width, $(\Delta  x)^2 = \int\rd x\, (x-\langle
x\rangle)^2 |E|^2/\int\rd x\, |E|^2$, panel (b), corresponding to Fig. \ref{FG3}.
\label{FG4} }
\end{figure}

The sum of the Fourier powers $|C_1|^2 +|C_2|^2 = 1 -
\frac{|\hat{v}_1|}{2}\sin(\Omega_0t)\sin(2\omega_0t)$ is not conserved since the
system (\ref{EQ4}) with $\delta k=0$ has the energy
\eqb
\mathcal{E} = \frac{1}{\omega_0^2}\left[\left|\frac{\rd C_1}{\rd t}\right|^2 +
\left|\frac{\rd C_2}{\rd t}\right|^2\right] + |C_1|^2 +|C_2|^2 +
2\mathrm{Re}(\hat{v}_1C_1C^*_2).
\label{EQ7}\eqe
For the solution (\ref{EQ6}) one can neglect the term
Re$\left(\hat{v}_1C_1C^*_2\right)=O(|\hat{v}_1|^2)$. Hence, there are oscillations
between the sum $|C_1|^2 +|C_2|^2$ and the analogous sum of the time derivatives of
$C_{1,2}$,  with the two frequencies $\omega_0\pm \Omega_0/2$,
corresponding to the lower and upper band-gap ends, and with the amplitude
$|\hat{v}_1|/4$. The oscillations reflect the transfer of the electromagnetic energy between the
electric and magnetic fields.  The Rabi oscillations can be traced also in the real space, where
they appear in the form of pulse average position oscillations, see Fig
\ref{FG4}(a). The pulse is also spreading (Fig. \ref{FG4}(b)) and propagating with
a small group velocity.

\begin{figure}[htb]
\centerline{\includegraphics[width=6cm]{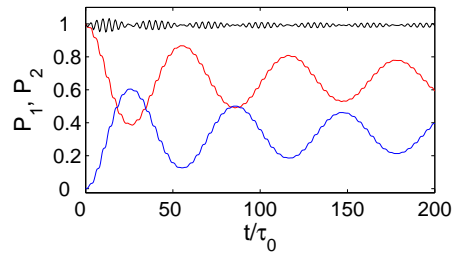}} \caption{(Color online) The
average Fourier powers $P_{1,2}$ of the numerical solution of Eq. (\ref{EQ1}) and
their sum (the upper line).  The difference from Fig. \ref{FG3} is in the pulse
width $\sigma = 6d$ which gives $\Delta k = 0.11k_B$, i.e. $\Delta\omega = 0.22
\omega_0 = 1.9\Omega_0$. }
\label{FG5}
\end{figure}

As one would expect, for shorter pulses than that of Fig. \ref{FG3}  the
oscillations show such features as the dephasing and the amplitude damping, since there is a
frequency detuning between  the resonant Fourier modes $k$ and $k-2k_B$ ($k\ne
k_B$), see Fig. \ref{FG2}. In the general case, the two resonant  modes in  Eq.
(\ref{EQ4}), i.e. $C_1(t) = C(k_B+\delta k,t)$ and  $C_2(t) = C(-k_B+\delta k,t)$,
have the following form (up to the $O(|\hat{v}_1|)$-terms)
\begin{eqnarray}
C_1 &=& e^{-i\overline{\omega} t}\left[\cos\left(\frac{\Omega t}{2}\right) -
i\frac{\gamma}{\sqrt{1+\gamma^2}}\sin\left(\frac{\Omega t}{2}\right)\right],
\nonumber\\
C_2 &= & e^{i\chi-i\overline{\omega}
t}\frac{1}{\sqrt{1+\gamma^2}}\sin\left(\frac{\Omega t}{2}\right),
\label{EQ8}\end{eqnarray}
where
\eqb
\overline{\omega} = \sqrt{\frac{\omega_1^2 +\omega_2^2}{2}},\; \Omega =
\overline{\omega}|\hat{v}_1|\sqrt{1 +\gamma^2},\; \gamma =
\frac{\omega_1^2-\omega_2^2}{\omega_1^2+\omega_2^2}\frac{1}{|\hat{v}_1|}.
\label{EQ9}\eqe
Since  the small $O(|\hat{v}_1|)$-terms are neglected, the total power is
conserved in Eq. (\ref{EQ8}) $|C_1|^2 +|C_2|^2 = 1$. Both the amplitude
of oscillations and the  frequency are defined by the ratio $\gamma \approx \frac{\delta
\omega}{\overline{\omega}|\hat{v}_1|}\approx \frac{\delta \omega}{\Omega_0}$ of the
frequency detuning to the band-gap width. Averaging over an interval of the size
$\Delta k$ and evaluating the integral by the stationary phase method,  one obtains
for $t\gg 1$:
\begin{eqnarray}
P_{1,2} &\equiv&\frac{1}{\Delta k}\int\limits_{\pm k_B-\Delta k/2}^{\pm k_B+\Delta
k/2}\rd
\lambda|C(\lambda,t)|^2 \nonumber\\
&=& \frac{1}{2}\left\{1 \pm\frac{\Omega_0}{\Delta
\omega}\sqrt{\frac{\pi}{\Omega_0t}}\cos(\Omega_0t +\pi/4)\right\},\quad
\label{EQ10}\end{eqnarray}
where $\Delta \omega = \omega(k_B+\Delta k/2) - \omega(-k_B+\Delta k /2)$. \
An example of  the Rabi oscillations
with a large detuning (i.e. for a short pulse) is given in Fig. \ref{FG5}. In this
case, the average position of the pulse and the pulse width show nearly linear
dependence on time,  the pulse  spreads  and propagates in the
lattice significantly.

In conclusion, an electromagnetic pulse propagation in a switched-on photonic
lattice with the central frequency lying in the forbidden gap is an analog of the
Rabi model, where the two-level media is the pulse, i.e. its two resonant Fourier
modes, while the photonic lattice serves as a classical monochromatic field. The
Rabi frequency is given by the band-gap width and the oscillations have all the
characteristic features as in the original Rabi case, such as the amplitude damping
due to the dephasing.   An additional feature is the high-frequency modulation of the
oscillations due to the energy transfer between the electric and magnetic fields of
the pulse. The effect may have  applications, for instance, as a base for an
all-optical trap for light pulses.

This work  has been supported by the  CAPES and FAPESP of Brasil. The author
acknowledges  the hospitality of  Gleb Wataghin Institute of Physics at UNICAMP in
Brazil, where he has benefited from illuminating discussions with L. E. Oliveira
and S. B. Cavalcanti.

\end{document}